\documentclass[12pt]{article}
\usepackage{latexsym}

\def\tr{{\rm tr}} 
 \def\cH{{\cal H}}
\def\ket#1{\mid~\!\!\!{#1}~\!\!\rangle}
\def\bra#1{\langle~\!\!{#1}~\!\!\!\mid}
\def\+-{\buildrel + \over -}

\def\QM{quantum mechanics }
\def\qm{quantum mechanics}
\def\Q{quantum }
\def\q{quantum}
\def\QMl{quantum-mechanical }

\def\cP{{\cal P}}
\def\cH{{\cal H}}

\def\${\enskip$}
\def\M{measurement }
\def\m{measurement}
\def\I{interpretation }
\def\i{interpretation}

\def\WF {wavefunction }
\def\wf{wavefunction}

\begin{document}

\begin{center} {\LARGE \bf Wavefunction
reality,\\ indeterminate properties and\\
degrees of presence} \vspace{.5cm}

\bf \large  Fedor Herbut\\
\end{center}

\noindent {\footnotesize \it Serbian Academy
of Sciences and Arts, Serbia, Belgrade, Knez
Mihajlova 35}

\vspace{0.5cm} \noindent \rule{13.6cm}{.4pt}

\noindent {\bf Abstract}\\

\indent The degree-of-presence (of the \Q system) concept,
accompanying that of the wavefunction-reality postulate, is introduced and
studied in two ways. To begin with, an incomplete exposition of the present author's views is  given. Subsequently, a short
historical and philosophical review of
answers to the question about the meaning of
indeterminate individual-system probabilities is presented from the literature. It is done  in the form of a carefully selected collage of quotations mostly with polemic comments by the present author and with further elaboration of his point of view. The advocated notion of 'degree of presence' generalizes the intuitively most easily acceptable idea of 'delocalization' in (roughly called) wavelike behavior of a \Q system.\\


\noindent \rule{13.6cm}{.4pt}

\vspace{0.5cm} \normalsize \rm
%1
%%%%%%%%%%%%%%%%%%%%%%%%%%%%%%%%%%%%%%%%
\noindent {\bf 1. Introduction}
%%%%%%%%%%%%%%%%%%%%%%%%%%%%%%%%%%%%%%%%

\vspace{0.3cm}

\normalsize \noindent

The assumption of a {\it realistic} interpretation of the wavefunction  has a long standing. The origin of my personal
interest for the subject lies in my recent
research. A successful run of \QMl insights in intricate experiments assuming reality of the wavefunction of individual quantum systems has been performed by the present author (Herbut and Vuji\v{c}i\'{c}, 1997; Herbut 2008a; Herbut 2010a). Individual-system probability as an important open question has remained lingering on.

\rule{2cm}{.4pt}

\indent \scriptsize {\it E-mail address:}
fedorh@sanu.ac.rs

\pagebreak

\normalsize
The last two mentioned studies of the present
author deal with intricate variations on the
{\it two-slit interference} experiment. As it
is well-known, one has in its description a
state vector (vector of norm one) \$\ket{\psi}\$ that consists of
two components expressed via state vectors
\$\ket{\psi}_i,\enskip i=u,l\$ as follows
$$\ket{\psi}=\alpha\ket{\psi}^u\enskip
+\enskip \beta\ket{\psi}^l,\quad
\alpha,\beta\in\mbox{\bf C}\quad
|\alpha|^2+|\beta|^2=1,\quad
\alpha\not=0\not=\beta.\eqno{(1)}$$ The
components mean passing the \textbf{u}pper slit \$u\$
and the \textbf{l}ower slit \$l\$ respectively, and
they are orthogonal
\$\bra{\psi}^u\ket{\psi}^l=0\$ due to
disjoint respective domains \$D^u\$ and
\$D^l\$ (around the slits) on which they are
non-zero.

Further, the square moduli of the amplitudes
\$|\alpha|^2\$ and \$|\beta|^2\$ are the
probabilities that the particle can be
found, in a so-called which-way measurement,
to come from \$D^u\$ or \$D^l\$ respectively.
One wonders {\it how one should understand
these probabilities} having in mind the individual-system case.\\

One can find the following quotation on individual-system probabilities in a book of {\bf Pauli} (1994, p. 43):
\begin{quote} {\bf Quote Pa}:  "...
incapable of further reduction ... a primary
fundamental notion of physics."
\end{quote}

{\bf Saunders} (1998, p. 20) makes a similar
remark:
\begin{quote} {\bf Quote Sa}:
"Probability cannot be reduced to anything
else: that is what I have been saying
throughout."
\end{quote}

{\bf Butterfield} (2002) speaks of
individual-system probability as follows:
\begin{quote} {\bf Quote Bu:} "[O]rthodoxy
says: the probability distribution ascribed
by the state is not for values possessed at
some time or other; but for values yielded as result (pointer-reading) on a measurement
apparatus. It is curious for the
interpretation of state to invoke something
so extrinsic to the system; and furthermore,
for a typical measurement on a typical
quantum system, something so vast compared
with the system, so variable from occasion to occasion, and so vague. Indeed, it is surely
more than curious: it is unsatisfactory,
unless you have some general operationalist
or instrumentalist view of theories."\\
\end{quote}

Butterfield speaks only of the manner how we, gross human beings, can take cognizance of
the individual-system probabilities. The proposed 'degree of presence' concept is an attempt to grasp their true nature.\\

The term 'degree-of-reality' appears in the article by Busch and Jaeger
(2010). Their idea differs from that of 'degree of presence' (see subsection 3.7 below for details). Sudbery (2011) gives a comprehensive discussion relevant for the individual-system probability concept (see subsection 3.8. below).

Since indeterminate properties can be viewed
also as dispositions, the nice review article
of M. Su\'{a}rez (2007) is also worthy of
reading. It reviews Margenau's latencies,
Heisenberg's potentialities, Maxwell's
propensitons, and the selective
propensities of the author.

I feel that the question of the {\it physical
meaning of individual-system quantum
probabilities} deserves additional
study.\\

In section 2 I'll introduce and explain the degree-of-presence concept. Section 3 is devoted to a collage of relevant quotes from other authors on indeterminate events or properties accompanied by a detailed comparison with the views of the present author, which are further elaborated where suitable.

In the collage of quotes, the first three authors (Aristotle, Reichenbach and Heisenberg) are mainly precursors of the degree-of-presence notion. The quoted views of the following five (Popper, Shimony, Mermin, Busch and Jaeger, and Sudbery) are inspirative for comparison and further elaboration of my standpoint.\\

%2.
%%%%%%%%%%%%%%%%%%%%%%%%%%%%%%%%%%%%%%%%
\noindent {\bf 2. The present author's view of individual-system probabilities
as degrees of presence}\\
%%%%%%%%%%%%%%%%%%%%%%%%%%%%%%%%%%%%%%%%

As it is well known, quantum probabilities are measured on
ensembles of equally prepared systems in the
well-known frequentist way. One cannot doubt
the reality of the ensembles, in particular,
not the measurement results obtained in them. But then it is hard to imagine how one can doubt the reality of the states of their constituents - of the individual quantum systems that make up the ensemble.

The present author bases his position on this argument, and he  is
partial to the wavefunction-reality point of
view. It is assumed throughout that the
wavefunction describes not only the corresponding ensemble,
but also the state of the individual systems
of which the ensemble consists.\\

%2.1
%%%%%%%%%%%%%%%%%%%%%%%%%%%%%%%%%%%%%%%%
\noindent {\bf 2.1 Gleason and degree of presence}\\
%%%%%%%%%%%%%%%%%%%%%%%%%%%%%%%%%%%%%%%%

It is one of my interpretational postulates of the \Q formalism that the \WF and everything that it implies and is observable has {\bf reality}, i. e., it exists 'out there' whether we observe it or not. The basic and most important implication of the \WF is probability: if \$E\$ is an event (or property or statement, mathematically a projector) and \$\ket{\Psi}\$ is a \WF (actually, a state vector, but we won't distinguish the abstract and the popular concrete representation form of this entity), then \$\bra{\Psi}
E\ket{\Psi}\$ is the probability of \$E\$ in \$\ket{\Psi}\$. We know, thanks to Gleason (1957), that also the converse is true, there is no other state, meaning the wider class of mixed-or-pure states, i. e., of density operators, that would give the same probabilities.

The way I understand this is that \$\ket{\Psi}\$ {\bf consists of} these numbers as of a kind of its {\bf observable pieces}
when \$E\$ is running over the entire quantum logic \$\cP(\cH)\$ (the set of all events - \textbf{p}rojectors - defined in the state space - \textbf{H}ilbert space - \$\cH\$ of the system). In this manner, in the view of the present author, reality of \$\ket{\Psi}\$ logically implies that of \$\bra{\Psi}E\ket{\Psi}\$. Since these
are numbers from the doubly closed
interval \$[0,1]\$, they can be called
{\bf 'degrees of presence'} of the \Q
system in the corresponding elements of
\$\cP(\cH)\$. Since the projectors can,
according to their physical meaning, be
events or properties or \Q statements,
one can also speak of degrees of presence
of the \Q system in these \Q entities.

Naturally, one may view 'degree of presence' as a semantic variation of 'probability'.  But making terms and endowing them with a physical meaning in the context of other physically meaningful notions (cf my comparison with Shimony's thoughts in subsection 3.5) seems to the present author to be the only way to learn to grasp \Q reality.

My understanding of 'degree of presence'
is shaped on the concrete example of {\bf delocalization}, where I find it easiest to imagine it. If standard \QM tells me that a particle passes coherently both slits in a two-slit interference experiment, and the probability of passing the upper slit is non-zero but small (\$0<|\alpha |^2\ll |\beta |^2\$ in (1)), this does not mean, in my physical picture of the situation, that the passage of the upper slit is less real than that of the lower one; it means that the particle is in a lesser amount present there. This picture stems, of course, from the classical wave passing two openings simultaneously. For the time being I cannot do with fewer classical notions. (Incidentally, as to 'interference', cf Herbut (1992), and more can be read on 'coherence' in  Herbut (2005a).)

The present author proposes to {\it extend} the
interpretation of {\it delocalization}, i. e., of {\it degrees
of localization}, to the
spectrum of an arbitrary observable;
particularly, to an arbitrary event or
property or  proposition and  call it 'degree
of presence'. More specifically, one may use
the terminological variations {\bf degree of occurrence} of
an event, or {\bf degree of possession} of a
property, or {\bf degree of truth} of a
proposition respectively.

Perhaps it is not superfluous to stress again that, according to the advocated point of view, a lesser degree of presence does not mean less reality. Reality is all or none. It is none only if \$\bra{\Psi}E\ket{\Psi}=0\$. A lesser degree of presence means that the system (the \WF of which we are concerned with) is to a lesser degree present in the entity in question (like in the case of delocalization).

A \WF is
both reality and {\bf knowledge}. A degree of presence that is smaller than \$1\$ is not a partial or incomplete subjective information on a more complete \Q reality 'out there', but instead, it is complete information on an objective and complete form of a \Q reality.\\

One should note that here, like in
relative-state Everett theory (Ecerett 1957
and Everett 1973), 'potential' or 'latent'
is, by definition, banned from the
interpretation. All parts, all components of
the wavefunction, are real or actual (I use this term as a synonym for real), in
general, but the system is in different degrees present in them.\\

Incidentally, I should point out that I strongly disagree with the still widespread view, originating with the main founder of \QM Niels Bohr (Petersen 1963) that the \QMl formalism has no other purpose than to provide us with numerical values for probabilities, which can then be tested experimentally. (Some 'rebels' against this prejudice call it "Shut up and calculate".) I believe that the \QMl formalism should teach us to {\bf comprehend \Q phenomena}  with as few prejudices from classical physics as possible. (Still, it might be considered a good thing that Bohr stopped large-scale inquiry into \Q insights, when he did it. It might have slowed down the rapid progress of the practical side of \QM that we have had up till now.)\\

\pagebreak

%2.2
%%%%%%%%%%%%%%%%%%%%%%%%%%%%%%%%%%%%%%%%
\noindent {\bf 2.2 Subsystems and reduced density operators}\\
%%%%%%%%%%%%%%%%%%%%%%%%%%%%%%%%%%%%%%%%

One must keep in mind the fact that the physical systems of nature always come entangled. We know from the \Q formalism that this means that states of subsystems are usually described by reduced density operators that are more general than \wf s. The standard argument goes as follows: $$\bra{\Psi}_{12}(A_1\otimes I_2)\ket{ \Psi}_{12}=\tr[(\ket{\Psi}_{12}\bra{\Psi}_{12}
)(A_1\otimes I_2)]=\tr(\rho_1A_1),\eqno{(2a)}$$ where $$\rho_1\equiv\tr_2(\ket{\Psi}_{12}\bra{\Psi}_{12}
)\eqno{(2b)}$$ ('$\tr_2$' denoting the trace over subsystem \$2\$) is the state (reduced density operator) of subsystem \$1\$.

Hence, probabilities of subsystem observables (like of \$A_1\$) are actually predictions of the \WF of the larger system (of system \$1+2\$ in the notation of relation (2a)), and the concept of 'degree of presence' refers necessarily also to the reduced density operators, i. e., to {\bf improper mixtures} (D'Espagnat 1976).\\

It should be pointed out that the 'degree of presence' concept does not apply to {\bf proper mixtures}, which are density operators just like the improper ones, but physically they contain incomplete knowledge or ignorance. It was stated also by Heisenberg that proper mixtures lack full objectivity (cf subsection 3.3).

As a speculation, I think that one might use the term 'believed degree of presence' in the \Q Bayesian way (cf Caves et al. 2002) for the individual-system probability implied by a proper mixture.\\

%2.3
%%%%%%%%%%%%%%%%%%%%%%%%%%%%%%%%%%%%%%%%
\noindent {\bf 2.3 Entanglement and relative states}\\
%%%%%%%%%%%%%%%%%%%%%%%%%%%%%%%%%%%%%%%%

There is a widespread dellusion that when one deals with a \WF \$\ket{\phi}\$ of a given physical system, then one may think of a world-\WF that factorizes tensorically $$\ket{\phi}\ket{\Phi}_{rest-of-the-world}.$$
For this to be valid the system under consideration would not be allowed to have interacted with any part of the rest of the world ever up till the present moment. This is so because entanglement is a kind of remnant of interaction, and it remains preserved in some form (cf Hulpke et al. 2006 and Herbut 2005b). Tensor factorization, on the other hand, means lack of entanglement.\\

One should call to mind that a \WF \$\ket{\phi}\$, as a rule, describes a system in a \Q experiment, and its state is obtained by preparation. As a consequence, the \WF is a \textbf{conditional state} in the collapse interpretation of \QM  (cf my unpublished study Herbut (2010b)).  It is a \textbf{relative state} in the no-collapse approach (Herbut 2011). I make use of the terms "approach" and "view" as of synonyms of "\i ".

The advocated view of \QM was called 'relative reality of unitarily evolving state' (RRUES by acronym) in my recently published study Herbut (2008a). I am partial to the no-collapse or relative-state interpretation of \QM following Everett (1957 and 1973), which rests on 'unitary \qm ', in which the Schr\"odinger equation - or equivalently the unitary evolution operator - rules all dynamics and collapse never happens. (More on 'unitary \qm ' in my comments on some thoughts of Sudbery in subsection 3.8 .)\\

%2.4
%%%%%%%%%%%%%%%%%%%%%%%%%%%%%%%%%%%%%%%%
\noindent {\bf 2.4 Standard 'actualization' and occurrence}\\
%%%%%%%%%%%%%%%%%%%%%%%%%%%%%%%%%%%%%%%%

{\bf Maximal degree of presence} has a special position in standard \qm . It is interpreted as actualization, collapse, objectification etc., or specifically as occurrence of an event, possession of a property, truth of a \Q proposition.  Let me present a first discussion of this notion as I see it.

To be concrete, I'll confine myself to events (projectors) \$E\$. If in a pure or improper mixed quantum state (density operator) \$\rho\$ an event satisfies \$\tr(\rho E)=1\$ (degree of presence equalling \$1\$), then and only then one says in standard (collapse-interpretation) \QM that the event occurs in this state.

The basic role of the concept of 'degree of presence', or rather of 'degree of occurrence' in the case of events at issue, is, in my opinion, to put all positive values of individual-system probabilities {\bf on equal footing} as far as  reality is concerned. Occurrence takes place in different degrees.

Quantum events come in pairs \$E,E^c\$, where \$E^c\equiv I-E\$ (\$I\$ being the identity operator) is the opposite event (the ortho-complementary projector). For all positive but non-maximal degrees of occurrence one has $$1>\tr( \rho E)> 0<\tr(\rho E^c)<1.$$

A kind of '\Q observer', if such existed, might take cognizance  - via \Q correlations - of such a pair of events simultaneously. But {\bf classical physics} knows only the extreme degrees of occurrence: an event occurs or it  does not (then the opposite event occurs).

{\bf Observation} of occurrence by human experimenters is performed with the help of {\bf classical apparatuses}. Hence we, gross classical beings, must have actualization
in the standard meaning, i. e., with maximal degree of presence.

But what is standard 'actualization' actually, one wonders. Of the many answers, let me just mention three: it is collapse of the \WF on the classical apparatus according to Bohr (and the Copenhagen \I Stapp(1972)); it is collapse in the consciousness of the sentient observer according to von Neumann (1955, last chapter); and finally, it is
being in one of the branches or in one of the component universes of the multiverse according to the Everettian approach (Everett 1957, Everett 1973, De Witt 2004).

I am attracted to the Everett approach that guides my present investigations. But I have doubts (see my discussion in subsection 3.8 and the second concluding remark in section 4.).\\

%3
%%%%%%%%%%%%%%%%%%%%%%%%%%%%%%%%%%%%%%%%
\noindent {\bf 3. Review and comparison}
%%%%%%%%%%%%%%%%%%%%%%%%%%%%%%%%%%%%%%%%

\vspace{0.3cm} \indent In this section a
historical and philosophical selection of
views on the notion of indeterminate
properties predicted by a wavefunction is
presented from the literature. The selection
is, of course, incomplete and subjective. The review is displayed in the form of a collage
of quotations accompanied by comments
to find out, if possible, where the views of the reviewed author differ from those of the present author.\\

Incidentally, there are five terms in the literature (that I know of) designating positive probability less than $1$ in application to individual
quantum systems: "indeterminacy",
"indefiniteness", "potentiality",   "propensity" and "latency"   .\\

\pagebreak

%3.1
%%%%%%%%%%%%%%%%%%%%%%%%%%%%%%%%%%%%%%%%
\noindent {\bf 3.1 Aristotle}\\
%%%%%%%%%%%%%%%%%%%%%%%%%%%%%%%%%%%%%%%%

\indent In a remarkable recent article {\bf
Sudbery} (2011 pp. 8,9) writes: \begin{quote}
{\bf Quote Su1:} ''Aristotle, in a famous
passage (Aristotle 1980), considered the
proposition "There will be a sea-battle
tomorrow". He argued that this proposition is
neither true nor false (otherwise we are
forced into fatalism). Thus he rejected the
law of excluded middle for future-tense
statements, implying that they obey a
many-valued logic. Modern logicians (Prior
1967) have considered the possibility of a
third truth-value in addition to "true" or
"false', namely u for "undetermined", for
future-tense statements. But, interestingly,
Aristotle admitted that the sea-battle might
be more or less likely to take place. This
suggests that the additional truth values
needed for future-tense statements are not
limited to 1, but can be any real number
between 0 and 1 and should be identified with
the probability that the statement will come
true. Turning this round gives us an {\it
objective form of probability} (italics by
F.H.) which applies to future events, or to
propositions in the future tense;..."
\end{quote}

Aristotle couldn't, of course, imagine that one day some people would think, in principle, about everything, including the battle he was concerned with, in terms of a change of a wave function governed by the Schr\"odinger equation. But nevertheless, his thoughts might be considered to be an early, perhaps the earliest,  precursors of the \QMl probabilities.\\

%3.2
%%%%%%%%%%%%%%%%%%%%%%%%%%%%%%%%%%%%%%%%
\noindent {\bf 3.2 The three-valued logic of Reichenbach}\\
%%%%%%%%%%%%%%%%%%%%%%%%%%%%%%%%%%%%%%%%

\indent {\bf Reichenbach}
(1965, paragraph 30), writing about how to
characterize statements on latent or
potential events, says:
\begin{quote}
{\bf Quote Re:} "... This is achieved with the introduction of a third truth value of
indeterminacy. The meaning of the term
"indeterminate" must be carefully
distinguished from the meaning of the term
"unknown". The latter term applies even to
two-valued statements, since the truth value
of a statement of ordinary logic can be
unknown; we then know, however, that the
statement is either true or false. The
principle of the {\it tertium non datur}, or
of the {\it excluded middle}, expressed in
this assertion, is one of the pillars of
traditional logic. If, on the other hand, we
have a third truth value of indeterminacy,
the {\it tertium non datur} is no longer a
valid formula; there is a tertium, a middle
value, represented by the logical status {\it indeterminate}." \end{quote}

Quantum mechanics does contain the "truth value indeterminacy", but, actually, one has more: something like a {\it degree of indeterminacy}, the subjective aspect of
'degree of presence' if the \WF  is
concerned.

Three-valued logic and probability goes back
to Aristotle (1980). Important contributions
are also in Lukasiewicz (1970), Reichenbach
(1949), and Sudbery (2011).\\

%3.3
%%%%%%%%%%%%%%%%%%%%%%%%%%%%%%%%%%%%%%%%
\noindent {\bf 3.3 Heisenberg}\\
%%%%%%%%%%%%%%%%%%%%%%%%%%%%%%%%%%%%%%%%

\indent {\bf Heisenberg}
(1958, p. 53) wrote: \begin{quote} {\bf Quote
He1:} "The probability function combines
objective and subjective elements. It
contains statements about possibilities or
better tendencies ("potentia" in Aristotelian
philosophy), and those statements are
completely objective, they do not depend on
any observer; and it contains statements
about our knowledge of the system, which of
course are subjective in so far as they may
be different for different observers. In
ideal cases the subjective element in the
probability function may be practically
negligible as compared with the objective
one. The physicists then speak of a "pure
case."" \end{quote}

Predictions of wavefunctions are 'objective
statements' in Heisenberg's wording. I cannot disagree with that. "Statements about our knowledge, that may be different for different observers" are implied by proper mixtures. They are much
discussed lately (see Herbut 2004 and the
references therein). They are almost completely
excluded in this study because the 'degree-of-presence' concept is not applicable to them.

The only point where Heisenberg apparently
differs from the advocated degree-of-presence
interpretation of probabilities implied by a \WF  is that he seems
to allow for a small subjectivity in the latter. This
can be understood in two ways. Firstly, what he means may be that the \WF
describes precisely and completely a model
state of a model system, and this differs to
some extent from the real state of the real
system. The choice of such models may be due to the need to lessen complexity and to
subjectivity.

Secondly, he might have been aware of the ubiquitous entanglement; Schr\"odinger (1935b and 1936) has brought them into focus. If a \WF is a conditional state (or a relative state, which is formally equivalent Herbut (2011)), it may be (formally) obtained from a proper mixture of a larger system containing ignorance. If any of these explanations is valid, Heisenberg's  wording is acceptable to me.

Also Busch and Jaeger (2010, p. 20) point out that the notion of 'unsharp reality'  began
with {\bf Heisenberg} (1959, p. 140), who
claimed:
\begin{quote} {\bf Quote He2:} "...
a course of events in itself is not
determined by necessity but that the
possibility or rather the 'tendency' towards
a course of events possesses itself a kind of
reality — a certain intermediate level of
reality midway between the massive reality of
matter and the mental reality of an idea or
picture..." ( translated from German by Busch
and Jaeger). \end{quote}

Then the two authors add (Busch and Jaeger,
{\it ibid.}, p. 20). "He (meaning Heisenberg
- F. H.) notes that this concept of
possibility is given in quantum theory in the form of probability."

I have the impression that more than half a century ago Heisenberg, one of the great pioneers of \qm , was struggling both with the idea of "kinds of reality" and with the relation between "necessity" and randomness. The two dilemmas are absent from my RRUES approach (cf the third passage in 2.3). There is one kind of reality, the \Q reality given by the \wf , but, when particular events (or properties etc.) are concerned, it appears in different degrees of presence of the system, and they are observable in ensembles in the frequentist way. Necessity, i. e., determinism is universally valid in the multiverse. Randomness applies to the branches or particular worlds (more details in subsection 3.8).

I cannot help disagreeing with Heisenberg's words
"a course of events in itself is not
determined by necessity but that the
possibility or rather the {\it 'tendency' towards} (italics by F. H.) ...". The degree-of-presence notion has no dynamical side to it. In this respect Heisenberg is a precursor of Popper and his propensity (cf subsection 3.4.), not of the dynamically completely void 'degree-of-presence' concept.

Further Busch and Jaeger go on: \begin{quote} {\bf Quote
(B+J)1:} "Heisenberg thus goes beyond the
usual interpretation of the pure quantum
state \$\ket{\psi}\$ as the catalogue of all
actual properties - those with probability
equal to one - of an individual system in
that he considers \$\ket{\psi}\$ as the
catalogue of the potentialities of all
possible (sharp) properties Q of the system,
quantified by the probabilities
\$psi(Q)=\bra{\psi}Q\ket{\psi}\$".
\end{quote}

In the terminology of Busch and Jaeger,
'sharp properties' are those represented by
projectors, whereas the ones for which
positive operators that are not projectors
stand are 'unsharp properties'. The latter, if
implied by a wavefunction, are always
non-maximal degrees of presence.

I understand the words of Busch and Jaeger so that Heisenberg may have been the first author, or at least the first important author, who highlighted indeterminate properties of individual \Q systems.\\

%3.4
%%%%%%%%%%%%%%%%%%%%%%%%%%%%%%%%%%%%%%%%
\noindent {\bf 3.4 Popper's propensity}\\
%%%%%%%%%%%%%%%%%%%%%%%%%%%%%%%%%%%%%%%%

\indent {\bf Popper} (1982, p. 159) wrote:
\begin{quote} {{\bf Quote Po1}: "According to
this picture...all properties of the physical
world are dispositional, and the real state
of a physical system, at any moment, may be
conceived as the sum total of its
dispositions - or its potentialities, or
possibilities, or propensities."}
\end{quote}

These words sound as a precursor of Gleason's theorem (cf 2.1). {\bf Selleri} and {\bf van der Merwe} (1991,
p. 1385), in the conclusion of their nice
essay on Popper, write: \begin{quote} {\bf
Quote S+vdM:} "His proposal concerning {\it
propensities} looks like an unescapable
fundamental ingredient for any future
realistic physics." \end{quote}

Popper's propensity is {\it tendency for
'actualization'}, and the latter is
'occurrence' in the classical sense. Two more
quotations make clear his position. On p. 351
(Popper 1983) he says: \begin{quote} {\bf
Quote Po2}: "[T]here is an analogy between
the idea of propensities and that of forces -
especially fields of forces. \dots both ideas
draw attention to {\it unobservable
dispositional properties of the physical
world}." \end{quote}

Further, on p. 395 {\it ibid.} he states:
\begin{quote} {\bf Quote Po3}: "I propose to
interpret the objective probability of a
single event as a measure of an objective
{\it propensity} - of the strength of the
tendency, inherent in the specified physical
situation , to realize the event - to make it
happen." \end{quote}

Thus, Popper's 'propensity' is the tendency
for standard actualization, and even with some dynamical attributes. This is the reason why I cannot accept it (and I cannot agree with Selleri and van der Merwe). I have begun to explain my view of actualization in subsection 2.4, and I'll give further elaboration in my comments on some ideas of Shimony in subsection 3.5.\\

Thinking along lines that Popper initiated
has moved on (Su\'{a}rez 2004 and others)
apparently with more mathematical and
philosophical sophistication. But, it seems
to the present author, that the
reality-of-wavefunction point of view
requires a simpler idea for the indisputably
primitive concept of individual-system
probability.\\

%3.5
%%%%%%%%%%%%%%%%%%%%%%%%%%%%%%%%%%%%%%%%
\noindent {\bf 3.5 Shimony}\\
%%%%%%%%%%%%%%%%%%%%%%%%%%%%%%%%%%%%%%%%

\indent {\bf Shimony} (1993, second passage
p. 179), in the context of Einstein's
world-view, discusses indefinite quantum
properties  as follows. \begin{quote} {{\bf
Quote Sh1}:  "The first of Einstein's theses
...., and the one which appears to stand
highest in his philosophical hierarchy, is
that physical things "claim a 'real
existence' independent of the perceiving
subject." This thesis {\it is} consistent
with all the conclusions which we drew from
an analysis of Bell's theorem, the relevant
experiments, and the formalism of \qm .
However, this thesis of physical realism,
when separated from the rest of Einstein's
theses, leaves open the character of the real
existence of physical things. The foregoing
analysis led to radical conclusions regarding
the character of physical existence: i. e.,
that there are {\it objective indefiniteness,
objective chance, and objective probability}
(italics by F.H.), in short, that there is a
modality of existence which has been
designated as potentiality. ..."}
\end{quote}

One can find additional elucidation of his
ideas in the following text ({\it ibid.}, p.
142, 2nd passage): \begin{quote} {\bf Quote
Sh2}: "The combination of indefiniteness of
value with definite probabilities  of
possible outcomes can be completely referred
to as {\it potentiality}, a term initially
suggested by Heisenberg (1958, p. 185). When
a physical variable which initially is only
potential acquires a definite value, it can
be said to be {\it actualized}. So far, the
only processes we have mentioned in which
potentialities are actualized are
measurements, but in a non-anthropocentric
view of physical theory the measurement
process is only a special case of the
interaction of systems, of special interest
to scientists because knowledge is thereby
obtained, but not fundamental from the
standpoint of physical reality itself...."
\end{quote}

Shimony's clear and illuminating text
highlights the concept of 'actualization'
in an inspirative, but standard way. He mentions \m , but does not (at least not in the above excerpts) really point to the great problem connected with \m . Namely, "in a non-anthropocentric
view of physical theory the measurement
process is only a special case of the
interaction of systems" - he says. But in a collapse \I of \QM this is not so because 'collapse' is not a 'special case' of dynamical evolution (cf von Neumann 1955, p. 351). This is known as the '\M paradox'.

It is a problem only in collapse \i s. In a no-collapse (relative-state) approach \M does not constitute a problem. There well defined \M interaction Busch et al. (1996) leads to a so-called pre\M state of object plus measuring instrument, in which the \M results have been copied to the measuring instrument state. There is no problem except that the theory leads to the parallel worlds about which I'll say more in my comments on some ideas of Sudbery (cf subsection 3.8).

Another follower of Einstein, Bell found the Everettian no-callapse approach "extravagant" (on account of the parallel universes, no doubt). Therefore, he  tried to comprehend '\m ' within collapse \qm . He wrote an article (Bell 1990) under the provocative title "Against Measurement". This led to a beautiful controversy.

Reaction came from Peierls (1991) and from Gottfried (1991). The latter's article had the title "Does Quantum Mechanics Carry the Seeds of Its Own Destruction?". (My answer would be: Not \qm , but perhaps its collapse interpretation.) Gottfried's reaction was criticized by Whitaker (2008). Also Mermin (2006) reacted to Bell's criticism of \m . His argument, in its turn, was critically examined by Ghirardi (2008).

The difficulties with the '\m ' concept in collapse \QM are a strong indication that one should take the Everettian approach into serious consideration. But one might retort that one thus barters one burden for another, perhaps ending up with a more serious one. One must find out which 'burden' is the lighter one, which can be disposed of. The \M problem is with us for almost a century.\\

Measurement is performing the 'actualization' that Shimony is concerned with. I'll present now a short exposition of how I see the {\bf connection between the three key concepts, entanglement, degree of presence, and actualization}.

Let \$\ket{\Psi}_{12}\$ be an arbitrary state vector of a composite system (a so-called bipartite state). It always has a bi-orthogonal expansion with real coefficients (so-called canonical Schmidt expansion, cf subsections 2.1 - 2.3 in Herbut (2007) for a review) $$\ket{\Psi}_{12}=
\sum_ir_i\ket{i}_1\ket{i}_2,\qquad\forall i,i':\quad
\bra{i}_1\ket{i'}_1=\bra{i}_2\ket{i'}_2=
\delta_{i,i'}.\eqno{(3)}$$ The sum is finite or infinite; the number of terms is called the Schmidt rank. It is uniquely determined by the bipartite state, and so are the expansion coefficients \$\{r_i:\forall i\}\$, though the basis \$\{\ket{i}_1:\forall i\}\$  in expansion (3) is not. A good quantitative measure of the {\it amount of entanglement} in \$\ket{\Psi}_{12}\$ is the so-called mutual information (cf Herbut 2005c) \$I\equiv -\sum_ir_ilnr_i\$.

Let the Schmidt rank, say \$N\$, be, for simplicity, finite \$1<N<\infty\$. Then the maximal value of the mutual information is \$I_{max}=lnN\$, and \$r_i=1/N,\enskip i=1,2,\dots,N\$. Thus, for large \$N\$, we obtain a large amount of entanglement if the degrees of presence of the events \$\ket{i}_1\bra{i}_1\$ in \$\ket{\Psi}_{12}\$, \$r_i\$,  are very small. Conversely, if some of the degrees of presence \$r_i\$ are large, the mutual information is small. In the opposite extreme case, when there is no entanglement (N=1), the degree of presence is \$1\$.

In this sense the amount of entanglement and the degree of presence of suitable subsystem events, 'correlations' and 'correlata' (see the next subsection), are kind of {\it complementary} to each other.

The opposite-subsystem events \$\ket{i}_1\bra{i}_1\$ and \$\ket{i}_2\bra{i}_2\$ in relation (3) not only have the same degree of presence \$r_i\$, but they are also the relative states with respect to each other (conditional states of each other in collapse \i ). One can say that they carry information about each other.

Viewing things in the way I have just suggested, there is nothing special about 'actualization', when \$N=1\$. Loosely speaking, \Q systems can 'know about one another' for any non-zero degree of presence.
We need 'actualization' in the standard sense because we are incarcerated in a classical world, where the degrees of presence of events are confined to zero and one. If we want to understand the \Q world, we must consider all possible \Q values of the degrees of presence on the same footing. In other words, we should understand 'actualization' as a synonym of 'reality'. It is then valid for all positive degrees of presence.\\

%3.6
%%%%%%%%%%%%%%%%%%%%%%%%%%%%%%%%%%%%%%%%
\noindent {\bf 3.6 Mermin}\\
%%%%%%%%%%%%%%%%%%%%%%%%%%%%%%%%%%%%%%%%

{\bf Mermin} (1998) writes (in section III
with the title "The puzzle of objective
probability"):
\begin{quote} {\bf Quote Me1}:
"[T]he fundamental role probability plays in
\QM has nothing to do with ignorance (unlike
in classical physics - remark by the present
author). The correlata - those properties we
would be ignorant of - have no physical
reality. There is nothing for us to be
ignorant of.... We lack an adequate
understanding of how probability or
correlation is to be understood ...
throughout this essay I shall treat
correlation and probability as primitive
concepts..."
\end{quote}

"The correlations not the correlata" has
become a well-known mantra of Mermin's
"Ithaca interpretation" program, which I
consider to be very promising and, above all, inspiring. A multipartite
\WF  usually contains entanglement, the most
\Q kind of correlation; it is real and it
represents a significant part of the reality
of the \wf .

To illustrate the important notion of
correlata, let me upgrade the above two-slit
experiment (cf the Introduction) into the one in the real
experiment of Kim {\it et al.} (2000), where
instead of slits one has two excited atoms
that emit in superposition, i. e., coherently,  by cascade
de-excitation a pair of photons (in opposite
directions):
$$\ket{\Psi}_{12}=(1/2)^{1/2}
\Big(\ket{\psi}^u_1
\ket{\psi}^u_2\enskip +\enskip
\ket{\psi}^l_1\ket{\psi}^l_2\Big),
\eqno{(4)}$$ where \$i=1,2\$ label
the two photons, and \$j=u,l\$ refer to the two atoms that have emitted the two photons (counterparts of the upper and the lower slit).

The subsystem events (projectors)
\$\ket{\psi}^u_1\bra{\psi}^u_1\$ and
\$\ket{\psi}^u_2\bra{\psi}^u_2\$ of the
opposite photons, cascade partners meaning
emission from the '\textbf{u}pper' atom, are
correlated in \$\ket{\Psi}_{12}\$: if
photon \$1\$ is emitted from the atom \$u\$,
then so is photon \$2\$ due to the cascade
de-excitation. They, and symmetrically
when one replaces \$u\$ by \$l\$, are the
{\it correlata}. Mermin is right claiming
that they "have no physical reality" if, in keeping with the standard view, one
restricts oneself to maximal degree of presence or,
speaking in the concrete terms of events,
when one defines as occurrence only
probability-one occurrence as in classical
physics and in the Copenhagen interpretation
of \QM (see Stapp 1972).

In the approach supported by the present
author one takes
into account all possible degrees of presence
from zero to one (in particular, all degrees of occurrence in the open interval \$(0,1)\$).
Therefore, in the view of the present author, the above correlata {\it do have reality}, but their degree of presence, i. e., the degree of the systems presence in the correlata is less than \$1\$; in particular, it is \$1/2\$.

Let me repeat that, to my mind, a lesser degree of presence does not mean less reality. Reality is all or none. A lesser degree means that the \Q system is 'there' in a smaller amount (like in delocalization).

Thus, Mermin and myself seem to share the belief in reality of the individual-system probabilities; though he, unlike myself, may not extend this to all probabilities. But we certainly differ in our views on the correlata. Mermin's words "There is nothing for us to be ignorant of" are not valid in this approach.\\

Perhaps Mermin's and my views on reality  differ more than one would suspect at first sight. Mermin determines the concept of correlations in terms of mean values of all observables (Hermitian operators) of the subsystems that are correlated (Mermin 1998, p. 754)) including, of course, projectors. He attributes reality to the correlations, and reality of the probabilities seems to be a consequence as far as they partake in correlations. Then, probabilities implied by improper mixtures might not be real for Mermin, though they are real for me (cf subsection 2.2).

In a preliminary study (Herbut 2008b, not published yet) it was shown that, following Mermin, one can restrict the observables used by Mermin to events, and {\it ipso facto} determine the correlations in terms of probabilities only. (One can essentially further restrict even the set of events utilized.) In this manner, {\bf one can derive the reality of correlations from that of probabilities}.

Let me put this clearly: In my view, on ground of Gleason's theorem, the individual-system probabilities derive their {\bf reality} directly from that stipulated for the \WF (cf 2.1). Correlations derive theirs from that of the probabilities, and so do the correlata if their probability is positive.

But, of course, I may have misunderstood Mermin. If he believes in the reality of the wave function, then it seems unlikely that he can doubt the reality of any observable part of it.\\

Perhaps surprisingly, Mermin and myself seem to agree up to a point in utilizing the no-collapse relative-state approach to understand actualization. He says (on p. 755) of Mermin (1998):
\begin{quote}
{\bf Quote Me2} "To the extent that "I" am describable by physics, which deals only with the correlations between me and the photomultipliers (phm), physics can only (correctly) assert that photomultiplier phm-n firing is perfectly correlated with my knowing that photomultiplier phm-n fired for either value of \$n\$. The question that physics does not answer is how it can be that I {\it know} that it is phm-1 and is {\it not} phm-2. This is indeed a problem. It is part of the problem of consciousness."
\end{quote}

Mermin's formulation "to the extent that "I" am describable by physics" shows reservation towards the no-collapse approach. I do not share this reservation. But still he allows the approach, and I agree with him about the perfect correlations. Instead of "I" in his text, one could have a classical measuring apparatus. Then, "to the extent that" \QM is extended to classical systems, in particular detectors, one has the famous paradox of \M in \qm . If, on the other hand, Mermin's "I" is replaced by "my friend", then we have the known paradox of Wigner 's friend (Wigner 1961).

An Everettian no-collapse approach, the point of view that is taken by me, treats all these cases on the same footing (see subsection 3.8). This approach is called the 'relative-state' one precisely because
every \WF is taken in relation to a well-defined subsystem and pure state of it (in Everett's original expositions) or any event on the subsystem (in my straightforward elaboration of Everett, Herbut 2011). This subsystem plus an event on it is a subject entity. It is a premise, in my view, of Mermin's text (or of an analogous text with the measuring apparatus or with Wigner's friend).

When Mermin is aware of seeing phm-1 and not any other value of \$n\$, one has a different subject entity: Mermin plus the event 'he is aware of phm-1'. In the relative-state approach this is not a "problem of consciousness" (or a problem of a classical measuring apparatus, or of Wigner's friend). It is just a consistent use of the \QMl formalism. (It need not be put aside as in Mermin's presentation.)

What has been said should, of course, be  distinguished from the problems like 'what is conscousness', 'what are classical systems', 'what is Wigner's friend when Wigner has not asked him yet what result he sees'. For me all "photomultiplier phm-n firings" are realities, but relative to different subject entities. This is why I call my understanding of the no-collapse relative-state approach 'relative reality of unitarily evolving states (RRUES)' (cf 2.3) as in my previous study Herbut (2008a).\\

%3.7
%%%%%%%%%%%%%%%%%%%%%%%%%%%%%%%%%%%%%%%%
\noindent {\bf 3.7 Busch and Jaeger}\\
%%%%%%%%%%%%%%%%%%%%%%%%%%%%%%%%%%%%%%%%

\indent \textbf{Busch and Jaeger} (2010
subsection 2.4 there) write:
\begin{quote}
{\bf Quote (B+J)2:} "We noted above that what is actual has the power to act. Similarly,
when a property is absent there is no power
to act. If this idea is extended to apply to
the intermediate mode of existence,
potentiality, one may say that an
indeterminate property has a quantifiable,
limited {\it degree of reality} (italics by
F. H.) that manifests itself in a limited
capacity — {\it potentiality} — to induce the associated measurement outcome. A
quantitative measure of the degree of reality and associated potentiality is given by the
quantum mechanical probability, which
provides the likelihood for an individual
outcome to occur in the event of
measurement."
\end{quote}

The authors of this excerpt use the term "degree of reality". For them the indefinite properties of individual-system \wf s are {\it not on the same footing} as the definite properties (maximal degree of presence). It is claimed that the former have limited reality, i. e., less than full reality, if I understand it correctly. Besides, I cannot agree with the dynamical ideas ascribed to indefinite properties like "power to act" or lack of it.

They write "limited capacity — {\it potentiality} — to induce the associated measurement outcome". From my point of view they fail to make a distinction between more
inherent attributes of the state of the individual system, such as the degrees of presence in case of a \wf , and the more external ones that come to the fore when we observe them (in the frequentist way, cf the first concluding remark in section 4.). But even the latter are, in my opinion, without any "power to act". It appears that  Busch and Jaeger  are under the influence of Popper's propensity, which I have criticized above.

Actualization comes about in \m , when the \Q world meets the classical one, and degrees of presence become converted, via  ensembles, into frequencies. Raising actualization to an exalted position in comparison with the degrees of presence that are less that \$1\$ when considering more inherent properties of the state of the individual system may result in a failure to grasp \Q essentials.\\

Another interesting passage from the article
of Busch and Jaeger ({\it ibid.} on p. 2)
reads:
\begin{quote}
{\bf Quote (B+J)3:} "On
a realist interpretation of Quantum Mechanics
as a complete theory, the referent of quantum
mechanical propositions is the individual
system. This would not only recognize the
possibility that such a philosophically
realist interpretation could in the end
enable the best description of the physical
world; it also has the potential benefit of
providing us with guidance in developing new,
appropriately adapted intuitions about
microphysical objects."
\end{quote}

This is a good answer to the possible objection to the
advocated approach claiming that
degree-of-presence is just another term for
probability. It is more than that. It means putting a
concept, individual-system probability, in
its proper place, in its proper conceptual
context. This may pave the way for a deeper understanding.\\

\pagebreak

%3.8
%%%%%%%%%%%%%%%%%%%%%%%%%%%%%%%%%%%%%%%%
\noindent {\bf 3.8 Sudbery}\\
%%%%%%%%%%%%%%%%%%%%%%%%%%%%%%%%%%%%%%%%

Now I present Sudbery's version of, what is called, the many-worlds or the multiverse view of universally valid \qm . Subsequently, I'll give my comments that will make clear to what extent I do agree with him.

On page 4  Sudbery (2011) one can read:
\begin{quote}
{\bf Quote Su2:} "{\bf External truth:} The truth about the universe is given by a state vector \$\ket{\Psi}_U\$ in a Hilbert space \$\cH_U\$ evolving according to the Schr\"odinger equation. If the Hilbert space can be factorised as \$\cH_U=\cH_S\otimes\cH_E\$, where \$\cH_S\$ contains states of an experiencing observer, then $$\ket{\Psi (t)}=\sum_n\ket{\eta_n} \ket{\Phi_n(t)},\eqno{(5)}$$
and all the states \$\ket{\eta_n}\$ for which \$\ket{\Phi_n(t)}\not= 0\$ describe experiences which actually occur at time t.
{\bf Internal truth} from the perspective \$\ket{\eta_n}\$: I actually have experience \$\eta_n\$ which tells me that the rest of the universe is in the state \$\ket{\Phi_n(t)}\$. This is an
objective fact; everybody I have talked to agrees with me."
\end{quote}

Essentially, we are back at the problem of Mermin (cf quote "Me2"), of Wigner's friend, of the \M paradox etc.

To begin with, I would prefer to include in
\$\ket{\eta_n}\$ (one possible version of) the entire classical world with all possible active or passive observers in it. Sudbery's insistence on subjective "experience" might go under the name 'many-minds' instead of 'many worlds' as done occasionally. I do not think that there is any important difference.

Sudbery does not mention decoherence, which
has been introduced implicitly by Zeh (1970) as the dynamical source for the expansion (5) (cf also Joos et al. (2003), especially Zeh's article there). Further, the terms in (5) are then considered as the parallel worlds.

The lhs of (5), representing the "external truth", is the state of the multiverse. It changes in a deterministic fashion according to the Schr\"odinger equation, and, I agree with Sudbery, that it is devoid of probabilities. But it is devoid of other attributes as well (see my 'doubt' below). The probabilities appear in the "internal
truth".

What I find disturbing is the fact that my own person (as well as yours, dear reader) is included in \$\ket{\eta_n}\$, and, in Sudbery's internal view, this is the case for just one value of \$n\$. But in Sudbery's external view there are many values (actually infinitely many) of \$n\$ all include me and all living creatures. This makes us a {\bf component not a subsystem} (like the infamous cat of Schr\"odinger (1935a), which appears as a live and a dead component). The subsystem is given by \$\cH_S\$ and it comprises all the components.

If this is all true, an ancestor of mine, when technology advances sufficiently, may, like in television stories, 'meet' or rather become one, with  another component of himself.

Disturbing or not, if one assumes universal validity of the \Q formalism, it leads to decomposition (5).\\

I am personally prepared to go along with the expounded Everettian universally-valid \QMl vision of the multiverse all the way up to relation (5). But here I have a serious {\bf doubt}.

Aware of a post-Everett standpoint of Wheeler (1977, p. 2), the very mentor and promoter of Everett's theory in the beginning (cf Byrne 2010 and Wheeler 1957), who, in turn, followed Wigner (1973, pp. 382-383), I share their opinion that the lhs of (5) {\bf has no physical meaning}.

The wave functions \$\{\ket{\Phi_n(t)}:\forall n\}\$ (cf (5)) from Sudbery's internal view are, in my opinion, legitimate extrapolations from our \Q experiments in the physical laboratory, but the lhs of (5), i. e., the \WF \$\ket{\Psi (t)}_U\$ of the multiverse ('universe' in Sudbery's terminology), is not. There is no external observer because \$U\$ is by definition everything. Then observables, and \m , and all the concepts of which \QM consists are not valid for this \wf . One obtains it as a "view from nowhere" (as Sudbery mentions). I think that no-collapse-oriented foundationally-minded physicists and philosophers are still wrestling with the idea of parallel universes (cf Saunders et al. 2010 and Carr 2007).\\

One might consider \$\ket{\Psi}_U\$, the lhs of in (5), as deprived of physical meaning, but still as a valid {\it mathematical 'envelope concept' or 'umbrella notion' to encompass all possible proper subsystem states}, which have full physical meaning.\\

For the reason of analogy, let me take resort to special relativity theory. The Minkowski space has itself no immediate physical meaning, but every one of its elements, points, is a possible event with full physical meaning. The space is just a mathematical 'envelope concept' enabling us to view all relevant possibilities.\\

Incidentally, in the Everettian expansion (5) \$\ket{\Phi_n(t)}\$ is a {\it relative state} taken relative to the {\it subject state}
\$\ket{\eta_n}\$. In the language of textbook \QM the former would be called a conditional state, that would come about if the condition, the event \$\ket{\eta_n}\bra{\eta_n}\$, occurred giving rise to the collapse of the lhs of (5) into one of the terms \$\ket{\eta_n} \ket{\Phi_n(t)}\$ on the rhs.\\

Another important passage in Sudbery's
article argues against a Bayesian (purely
subjectivistic, Caves {\it et al.} 2002)
interpretation of probability. It reads as
follows ( {\it ibid.} p. 8).
\begin{quote}
{\bf Quote Su3:} "... our experience in a
quantum-mechanical world seems to require a
description in terms of objective chance.
Things happen randomly, but with definite
probabilities that cannot be reduced to our
beliefs. The value of the half-life of
uranium 238 is a fact about the world, not a
mere consequence of someone's belief."
\end{quote}

The present author agrees with Sudbery completely concerning purely subjective individual-system  \Q probabilities.\\

%3.9
%%%%%%%%%%%%%%%%%%%%%%%%%%%%%%%%%%%%%%%%
\noindent
{\bf 3.9. Einstein's boxes and delocalization}\\
%%%%%%%%%%%%%%%%%%%%%%%%%%%%%%%%%%%%%%%%

\indent
Let me first mention the famous Einstein boxes (Norsen 2005; sometimes called De Broglie boxes). A wavefunction \$\ket{\psi}_p\$ describes a single \textbf{p}article homogeneously delocalized within a box. The box is then partitioned into two; one of the boxes thus obtained is brought to London and the other is taken to Tokyo. When two observers, one in London and one in Tokyo, open the boxes, only one of them will find the (whole) particle.

Nevertheless, \QM tells us that, before opening, the particle was not in one of the boxes. The wavefunction describing the particle delocalized in the two boxes implies coherence between its two one-box components, i. e., these are capable of interference. It would be impossible if the particle were in one of the boxes. (Then the particle would be described by a proper mixture.) Unfortunately, the London-and-Tokyo observers are unable to perform an interference experiment. But in the physically isomorphic case of a two-slit experiment (cf relation (1)) this can be done (at least in an ensemble).

We easily picture the mentioned opening of the boxes in the collapse \i . To describe it in the no-collapse, relative-state approach, let us assume, for simplicity, that the \textbf{o}bservers in \textbf{L}ondon and \textbf{T}okyo are also in a pure state \$\ket{\phi}_o^{LT}\$. Then, before opening the boxes we have the composite state \$\ket{\phi}_o^{LT}\ket{\psi}_p^{LT}\$, where \$\ket{\psi}_p^{LT}\$ is the state of the particle delocalized in the London and the Tokyo boxes.

The act of opening the boxes is, in principle, a dynamical evolution that converts the last mentioned composite state into $$(1/2)^{1/2}\Big(\ket{\phi}_o^L\ket{\psi}_p^L+ \ket{\phi}_o^T\ket{\psi}_p^T\Big),\eqno{(6)}$$ where the terms correspond to finding the particle in the London or the Tokyo box respectively. Note that the coherence is not destroyed; it is only elevated to the composite system. The particle is now in the proper mixture \$(1/2)\Big(\ket{\psi}_p^L\bra{\psi}_p^L+
\ket{\psi}_p^T\bra{\psi}_p^T\Big)\$. Needless to say that 'finding the particle in the London box', e. g., is the relative state  relative to the subject state \$\ket{\phi}_o^L\$ in the composite state (6).

If we choose to view the state after opening the boxes via the all-encompassing decomposition (5) that we have borrowed from Sudbery, then we must join the state \$\ket{\chi}_r\$ of the \textbf{r}est of the world in case of opening. But we must also allow for the possibility of not opening the boxes, because a decision on part of the observers can be made or not (in a parallel world it is not made), just like it is in the case of \M results. Besides, there may be also other world components.

Thus, instead of (6) we obtain
$$\alpha\Big(\ket{\phi}_o^L\ket{\psi}_p^L +\ket{\phi}_o^T\ket{\psi}_p^T)\Big)
\ket{\chi}_r+\beta\Big(\ket{\phi}_o^{LT}
\ket{\psi}_p^{LT}\ket{\chi}'_r\Big)+\dots ,
\eqno{(7)}$$
where \$\ket{\phi}_o^L\$ and  \$\ket{\phi}_o^T\$ are now the states of  Sudbery's "experiencing" observers who find the particle in the London or the Tokyo box respectively, and the product of the other two \wf s applies to Sudbery's 'rest of the world'.

It is assumed that 'the rest of the world' is not the same when the boxes are opened and when they are not. This makes the composite state (6) the relative state with respect to the subject state \$\ket{\chi}_r\$ in the larger composite state (7), etc. (This state has to be orthogonal to all the other states of the rest of the world in (7).)\\

%4.
%%%%%%%%%%%%%%%%%%%%%%%%%%%%%%%%%%%%%%%%
\noindent {\bf 4. Concluding remarks}\\
%%%%%%%%%%%%%%%%%%%%%%%%%%%%%%%%%%%%%%%%

One might say roughly that viewing the individual-system probabilities as degrees of presence is, following the German philosopher Immanuel Kant, Ding an Sich (thing for itself), and that measuring the probabilities in ensembles in the frequentist way, when we take cognisance of the their values, is Ding f\"ur Uns (thing for us). Strictly speaking, only the very existence of a \Q system under consideration, its reality,  is 'a thing for itself'. Our \Q knowledge \$\ket{\Phi_n(t)}\$ (cf relation (5)) about it, though it is a maximal \Q reality, it is relative reality, taken with respect to the state \$\ket{\eta_n}\$ of our mind, or of the measuring instrument etc. Even if we do not take (5) to be a physically meaningful decomposition of the \WF of everything, the lhs can be the \WF of another subject entity, e. g., of 'Wigner' when he considers his 'friend'. 'Relative reality' itself is a step towards the subject entity, and towards the 'thing for us'. The frequentist \M is only the last step.\\

At the time of writing this article I have a fundamental {\bf dilemma}. Namely, I believe that Occam's razor is a good principle for science. Applying it to the Everettian approach, we have a minimum of assumptions at the input, but a maximum of hardly acceptable   final outcomes (the parallel worlds) at the output. One wonders if one should not, perhaps, apply Occam's razor partly also to the output.

If I am allowed to speculate, it seeems to me that this can be done by postulating that each system has an 'individuality beyond \qm ' (not a \Q property). Then it won't branch into the different worlds and the system will remain a subsystem of the universe, not a component.

One way to achieve this might be to take  suitable hidden variables. The most elaborate theory along these lines utilizes the Bohmian (non\q ) positions of all particles (Bohm and Hiley 1993). (No two particles can be at the same place. Hence, their non-quantum position is their 'individuality beyond \qm '.)

Relation (5) is still valid, perhaps with only mathematical meaning. No more than  one term (one branch) is real in the classical sense; the one into which the 'individuality beyond \qm ', together with those of the rest of the classical world, 'falls' in a random way. The other terms are, at the instant at issue, 'unrealized former possibilities'. But they still represent \Q reality, and they may partake in forming a future state.\\

%Ref
%%%%%%%%%%%%%%%%%%%%%%%%%%%%%%%%%%%%
{\bf \noindent References}\\
%%%%%%%%%%%%%%%%%%%%%%%%%%%%%%%%%%%%

\noindent Aristotle (1980). {\it De
Interpretatione}. Peripatetic Press,
Grinnell, Iowa.

\setlength{\parindent}{3ex} Chapter 9.

\noindent Bell, J. S. (1990). Against 'Measurement'. {\it Physics World}, August,
33-40.

\noindent Bohm, D. and Hiley, B. J. (1993). {\it The Undivided Universe. An Ontological

\setlength{\parindent}{3ex}Interpretation of Quantum Theory}. Routledge, London.

\noindent Busch, P., Lahti, P. J. and
Mittelstaedt P. (1996). {\it The Quantum
Theory of

\setlength{\parindent}{3ex}Measurement}. 2nd
edition. Lecture Notes in Physics. Springer,
Berlin.

\noindent Busch, P. and  Jaeger, G. (2010).
Unsharp Quantum Reality. {\it Foundations of

\setlength{\parindent}{3ex}Physics}, {\bf
40}, 1341-1367; arXiv:1005.0604v2.

\noindent Butterfield, J. (2002). Some Worlds
of Quantum Theory. In {\it Quantum Me-

\setlength{\parindent}{3ex} chanics}.
Scientific Perspectives on Divine Action.
Russell R. {\it et al.} eds.

\setlength{\parindent}{3ex} Vol. 5, 111-140.
Rome: Vatican Observatory Publications, 2;
arXiv:

\setlength{\parindent}{3ex} 0105052.

\noindent Byrne, P. (2010). {\it The Many Worlds of Hugh Everett III}. Oxford University

\setlength{\parindent}{3ex}Press, Oxford.

\noindent Carr, B. (2007) editor. {\it Universe or Multiverse?} Cambridge University Press,

\setlength{\parindent}{3ex}Cambridge.

\noindent Caves, C. M., Fuchs, C. A. and
Schack, R. (2002). Quantum Probabilities

\setlength{\parindent}{3ex}as Bayesian
Probabilities. {\it Physical Review A}, {\bf
65}, 022305 (6 pages);

\setlength{\parindent}{3ex}
arXiv:quant-ph/0106133v2.

\noindent D'Espagnat, B. (1976). {\it Conceptual
Foundations of Quantum Mechanics}.

\setlength{\parindent}{3ex}Second edition. W. A. Benjamin, Reading, Massachusetts, subsection 7.2

\noindent De Witt, B. and Graham, N. editors.
(1973). {\it The Many-Worlds Interpreta-

\setlength{\parindent}{3ex} tion of Quantum
Mechanics}. Princeton University Press,
Princeton.

\noindent De Witt, B. S. (2004). The Everett Interpretation of Quantum Mechanics.

\setlength{\parindent}{3ex}In {\it Science and Ultimate Reality}. Editors J. D. Barrow et al. Cambridge

\setlength{\parindent}{3ex}University Press, Cambridge. 167-198.

\noindent Everett, H. (1957). "Relative
State" Formulation of Quantum Mechanics.
{\it

\setlength{\parindent}{3ex}Reviews of Modern
Physics}, {\bf 29},  454-462. Reprinted in
De Witt and Gra-

\setlength{\parindent}{3ex}ham (1973),
141-150.

\noindent Everett, H. (1973). The Theory of
the Universal Wave Function.

\setlength{\parindent}{3ex}In De Witt B. and
Graham N. (1973), 1-140.

\noindent Ghirardi, G.-C. (2008). Reconsidering Mermin's "In Praise of Measurement".

\setlength{\parindent}{3ex}ArXiv:0806.0647

\noindent Gleason, A. M. (1957). Measures on
the Closed Subspaces of a Hilbert Space.
{\it

\setlength{\parindent}{3ex}Journal of
Mathematics and Mechanics}, {\bf 6},
885-893.

\noindent Gottfried, K. (1991). Does Quantum Mechanics Carry the Seeds of its Own

\setlength{\parindent}{3ex}Destruction? {\it Physics World}, October,
34-40.

\noindent Heisenberg, W. (1958). {\it Physics
and Philosophy}. Harper and Brothers Publi-

\setlength{\parindent}{3ex}cation, New York.

\noindent Heisenberg, W. (1959). The
Discovery of Planck and the Philosophic
Prob-

\setlength{\parindent}{3ex} lems of Atomic
Physics. (In German) {\it Universitas}, {\bf
14}, 135-153.

\noindent Herbut, F. (1992). Quantum Interference Viewed in the Framework of Pro-

\setlength{\parindent}{3ex}bability Theory. {\it American Journal of Physics}, {\bf 60}, 146-150.

\noindent Herbut, F. and Vuji\v{c}i\'{c} M.
(1997). First-Quantisation
Quantum-Mechanical

\setlength{\parindent}{3ex}Insight into the
Hong-Ou-Mandel Two-Photon Interferometer with
Pola-

\setlength{\parindent}{3ex}rizers and its
Role as a Quantum Eraser. {\it Physical
Review}, {\bf A 56}, 1-5.

\noindent Herbut, F. (2004). On Compatibility
and Improvement of Different Quantum

\setlength{\parindent}{3ex}State Assignments.
{\it Journal of  Physics A: Mathematical and
General}, {\bf 37},

\setlength{\parindent}{3ex}1-8;
arXiv:quant-ph/04051.

\noindent Herbut, F. (2005a). A Quantum Measure of Coherence and Incompatibility. {\it

\setlength{\parindent}{3ex}Journal of Physics}, {\bf A 38}, 1046-1064.

\noindent Herbut, F. (2005b). Can We Believe in a Purely Unitary Quantum Dynamics? ArXiv:quant-ph/0507218

\noindent Herbut, F. (2005c). Mutual Information of Bipartite States and Quantum

\setlength{\parindent}{3ex}Discord in Terms of Coherence Information.
{\it International Journal of

\setlength{\parindent}{3ex}Quantum Information}, {\bf 3}, 691-728.

\noindent Herbut, F.(2007). Quantum
Probability Law from 'Environment-Assisted

\setlength{\parindent}{3ex}Invariance'
 in Terms of Pure-State
Twin Unitaries. {\it Journal of Physics A:

\setlength{\parindent}{3ex}Mathematical and
Theoretical}, {\bf 40}, 5949-5971; arXiv:
quant-ph/0611220.

\noindent Herbut, F. (2008a). On EPR-type
Entanglement in the Experiments of Scully

\setlength{\parindent}{3ex}et al. I. The
Micromaser Case and Delayed-Choice Quantum
Erasure. {\it

\setlength{\parindent}{3ex}Foundation of
Physics}, {\bf 38}, 1046-1064; arXiv:
quant-ph/0808.3176 .

\noindent Herbut, F. (2008b). Quantum Correlations in Multipartite States. Study

\setlength{\parindent}{3ex}Based on the Wootters-Mermin Theorem. ArXiv:0811.3674

\noindent Herbut, F. (2010a). On EPR-type
Entanglement in the Experiments of Scully

\setlength{\parindent}{3ex}et al. II. Insight
in the Real Random Delayed-Choice Erasure
Experiment. {\it

\setlength{\parindent}{3ex}Foundations of
Physics}, {\bf 40}, 301-312; arXiv:
quant-ph/0808.3177.

\noindent Herbut, F. (2010b). A Theory of Quantum Preparation. ArXiv:1005.1348

\noindent Herbut, F. (2011). How Can the No-Collapse and the Collapse Interpreta-

\setlength{\parindent}{3ex}tions of Quantum Mechanics Give the Same Description? ArXiv:1108.4175

\noindent Hulpke et al. (2006). Hulpke, F., Poulsen, U. F., Sanpera, A., Sen(De),

\setlength{\parindent}{3ex}A., Sen, U., and Lewenstein, M.
Unitarity as Preservation of Entropy and

\setlength{\parindent}{3ex}Entanglement in Quantum Systems. {\it Foundations of Physics}, {\bf 36}, 477-

\setlength{\parindent}{3ex}499.

\noindent Joos, E. et al. editors (2003).
{\it Decoherence and the Appearance of a Classical

\setlength{\parindent}{3ex}World in Quantum Theory}.
Springer, Berlin.

\noindent Kim, Y.-H., Yu, R., Kulik, S. P.,
Shih, Y. and Scully, M. O. (2000). Delayed

\setlength{\parindent}{3ex}"Choice" Quantum
Eraser. {\it Physical Review Letters}, {\bf
84}, 1-5.

\noindent Lukasiewicz, J. (1970). Logical
Foundations of Probability Theory. In {\it
Se-

\setlength{\parindent}{3ex} lected Works}
edited by L. Borkowski. North-Holland,
Amsterdam, 16–63.

\setlength{\parindent}{3ex}(First published
in Cracow 1913).

\noindent Mermin, N. D. (1998). What is
Quantum Mechanics Trying to Tell us? {\it
Ame-

\setlength{\parindent}{3ex}rican Journal of
Physics}, {\bf 66}, 753-767; arXiv:
quant-ph/9801057v2.

\noindent Mermin, N. D. (2006). In Praise of Measurement. {\it Quantum Information

\setlength{\parindent}{3ex}Processing} {\bf 5}, 239-260; ArXiv:quant-ph/0612216

\noindent Norsen, T. (2005). Einstein's
Boxes. {\it American Journal of Physics},
{\bf 73}, 164-

\setlength{\parindent}{3ex}176.

\noindent Pauli, W. (1994). Probability and
Physics. In {\it Writings on Physics and
Phi-

\setlength{\parindent}{3ex}losophy}.
Springer-Verlag, New York 43-48.

\noindent Peierls, R. E. (1991). In Defense
of "Measurement". {\it Physics World},
January,

\setlength{\parindent}{3ex}19-20.

\noindent Petersen, A. (1963). The Philosophy of Niels
Bohr. {\it The Bulletin of the

\setlength{\parindent}{3ex}Atomic
Scientists}. September , p. 8.

\noindent Popper, K. R. (1982). {\it Quantum
Theory and the Schism in Physics}. Hutchin-

\setlength{\parindent}{3ex}son, London.

\noindent Popper, K. R. (1983). {\it Realism
and the Aim of Science}. Hutchinson, London.

\noindent Prior, A. (1967). Past, Present and
Future. Clarendon Press, Oxford.

\noindent Reichenbach, H. (1949). {\it The
Theory of Probability}. University of
California

\setlength{\parindent}{3ex}Press.

\noindent Reichenbach, H. (1965). {\it
Philosophic Foundations of Quantum
Mechanics}.

\setlength{\parindent}{3ex}University of
California Press, Berkeley. (First published
in 1944).

\noindent Saunders, S. (1998). Time, Quantum
Mechanics, and Probability. {\it Synthese},
{\bf

\setlength{\parindent}{3ex}114}, 373-404;
quant-ph/0111047.114 .

\noindent Saunders, S. et al. editors (2010). {\it Many Worlds?} Oxford University Press,

\setlength{\parindent}{3ex}Oxford.

\noindent Schr\"oddinger, E. (1935a). The Present Situation in Quantum Mechanics. A

\setlength{\parindent}{3ex}Translation of Schr\"odinger's "Cat Paradox" Paper. In Wheeler and Zurek

\setlength{\parindent}{3ex}(1983) 152-167.

\noindent Schr\"{o}dinger, E. (1935b). Discussion of Probability Relations Between Sepa-

\setlength{\parindent}{3ex}rated Systems. Proc. Cambridge
Phil. Soc. {\bf 31}, 555-563.

\noindent Schr\"{o}dinger, E. (1936). Probability Relations Between Separated Systems.

\setlength{\parindent}{3ex}Proc. Cambridge Phil. Soc. {\bf 32}, 446-452.

\noindent Selleri, F. and van der Merwe, A.
(1991). Karl Popper at Ninety: Highlights

\setlength{\parindent}{3ex}of a Lifelong
Intellectual Quest. {\it Foundations of
Physics}, {\bf 21}, 1375-1386.

\noindent Shimony, A. (1993). {\it Search for
a Naturalistic World View}. Vol. II. Cam-

\setlength{\parindent}{3ex}bridge University
Press, Cambridge.

\noindent Stapp, H.P. (1972). The Copenhagen
Interpretation. {\it American Journal of

\setlength{\parindent}{3ex}Physics}, {\bf
40}, 1098-1116.

\noindent Su\'{a}rez, M. (2004). On Quantum
Propensities; 2 Arguments Revisited. {\it
Er-

\setlength{\parindent}{3ex}kenntniss}, {\bf
61}, 1-6.

\noindent Su\'{a}rez, M. (2007). Quantum
Propensities. {\it Studies in History and
Philoso-

\setlength{\parindent}{3ex}phy of Modern
Physics}, {\bf 38}, 418-438.

\noindent Sudbery, A. (2011). Philosophical
Lessons of Entanglement. Talk given at

\setlength{\parindent}{3ex}75 Years of
Quantum Entanglement Kolkata, India. 10
January 2011.

\setlength{\parindent}{3ex}ArXiv:1103.4318v1.

\noindent Von Neumann, J. (1955) {\it Mathematical Foundations of Quantum Mechanics}.

\setlength{\parindent}{3ex}Princeton University Press, Princeton.

\noindent Wheeler, J. A. (1957). Assessment of Everett's "Relative State" Formulation

of Quantum Theory. {\it Reviews of Modern Physics}, {\bf 29}, 463-465.

\noindent Wheeler, J. A. (1977). Include the Observer in the Wave Function?  In {\it

\setlength{\parindent}{3ex}Quantum Mechanics, a Half Century Later} 1-18. Editors J. L. Lopes and

\setlength{\parindent}{3ex}M. Paty. D. Reidel Publishing Company, Dordrecht, Holland.

\noindent Wheeler, J. A. and Zurek, W. H. editors (1983). {\it Quantum Theory and Mea-

\setlength{\parindent}{3ex}surement}. Princeton University Press. Princeton.

\noindent Whitaker, M. A. B. (2008). {\it
Foundations of Physics}, {\bf 38}, 436-447.

\noindent Wigner, E. (1961). Remarks on the Mind-Body Question. In Wheeler and

\setlength{\parindent}{3ex}Zurek  (1983), 168-181.

\noindent Wigner, E. P. (1973). Epistemological Perspective on Quantum Theory.  In {\it

\setlength{\parindent}{3ex}Contemporary Research in the Foundations and the Philosophy of Quan-

\setlength{\parindent}{3ex}tum Theory}. Editor C. A. Hooker. Reidel, Dordrecht, Holland, 369-385.

\noindent Zeh, H. D. (1970). On the Interpretation of
Measurement in Quantum The-

\setlength{\parindent}{3ex}ory. {\it
Foundations of Physics}, {\bf 1},  69-76.

\end{document}